\documentclass{ws-ijmpa}

\begin{document}

\markboth{IVAN SCHMIDT}{Antishadowing}

\catchline{}{}{}{}{}

\title{SHADOWING AND ANTISHADOWING IN NEUTRINO DEEP INELASTIC
SCATTERING}

\author{IVAN SCHMIDT}
\address{Departamento de F\'\i sica,
Universidad T\'ecnica Federico Santa Mar\'\i a, \\ Casilla 110-V,
Valpara\'\i so, Chile \\ e-mail: ischmidt@fis.utfsm.cl}



\maketitle


\begin{abstract}
The coherence of multiscattering quark nuclear processes leads to
shadowing and antishadowing of the electromagnetic nuclear
structure functions in agreement with measurements.  This picture
leads to substantially different antishadowing for charged and
neutral current processes, particularly in anti-neutrino
reactions, thus affecting the extraction of the weak-mixing angle
$\sin^2\theta_W$.
\end{abstract}

\keywords{Shadowing; neutrino.}


\section{Introduction}

The weak-mixing angle $\sin^2\theta_W$ is an essential parameter
in the standard model of electroweak interactions. Until recently,
a consistent value was obtained from all electroweak
observables~\cite{Abbaneo:2001ix}. However, the NuTeV
Collaboration~\cite{Zeller:2001hh} has determined a value for
$\sin^2\theta_W$ from measurements of the ratio of charged and
neutral current deep inelastic neutrino--nucleus and
anti-neutrino--nucleus scattering in iron targets which has a $3
\sigma$ deviation with respect to the fit of the standard model
predictions from other electroweak
measurements~\cite{Abbaneo:2001ix}. Although the NuTeV analysis
takes into account many sources of systematic errors, there still
remains the question of whether the reported deviation could be
accounted for by QCD effects. Here we shall investigate whether
the anomalous NuTeV result for $\sin^2\theta_W$ could be due to
the different behavior of leading-twist nuclear shadowing and
antishadowing effects for charged and neutral currents~\cite{BSY}.

\begin{figure}[htb]
\begin{center}
\leavevmode {\epsfysize=6cm \epsffile{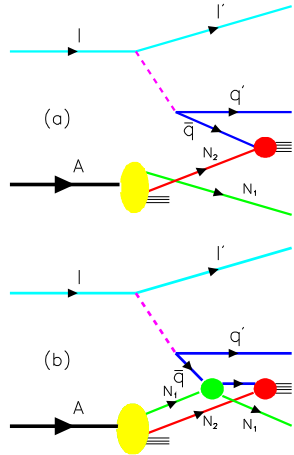}}
\end{center}
\caption[*]{\baselineskip 13pt The one-step and two-step processes
in DIS on a nucleus.  If the scattering on nucleon $N_1$ is via
pomeron exchange, the one-step and two-step amplitudes are
opposite in phase, thus diminishing the $\bar q$ flux reaching
$N_2.$ This causes shadowing of the charged and neutral current
nuclear structure functions. \label{bsy1f1}}
\end{figure}

The physics of the nuclear shadowing in deep inelastic scattering
can be most easily understood in the laboratory frame using the
Glauber-Gribov picture.  The virtual photon, $W$ or $Z^0$,
produces a quark-antiquark color-dipole pair which can interact
diffractively or inelastically on the nucleons in the nucleus. The
destructive and constructive interference of diffractive
amplitudes from Regge exchanges on the upstream nucleons then
causes shadowing and antishadowing of the virtual photon
interactions on the back-face nucleons. The coherence between
processes which occur on different nucleons at separation $L_A$
requires small Bjorken $x_{B}:$ $1/M x_B = {2\nu/ Q^2}  \ge L_A .$
An example of the interference of one- and two-step processes in
deep inelastic lepton-nucleus scattering is illustrated in
Fig.~\ref{bsy1f1}. In the case where the diffractive amplitude on
$N_1$ is imaginary, the two-step process has the phase $i \times i
= -1 $ relative to the one-step amplitude, producing destructive
interference (the second factor of $i$ arises from integration
over the quasi-real intermediate state.)  In the case where the
diffractive amplitude on $N_1$ is due to $C=+$ Reggeon exchange
with intercept $\alpha_R(0) = 1/2$, for example, the phase of the
two-step amplitude is ${1\over \sqrt 2}(1-i) \times i = {1\over
\sqrt 2} (i+1)$ relative to the one-step amplitude, thus producing
constructive interference and antishadowing. Due to the different
energy behavior, this also indicates that shadowing will be
dominant at very small x values, where the pomeron is the most
important Regge exchange, while antishadowing will appear at a bit
larger x values.

\vspace{0.3cm}

\section{Parameterizations of quark-nucleon scattering}

We shall assume that the high-energy antiquark-nucleon scattering
amplitude $T_{\bar{q}N}$ has the Regge and analytic behavior
characteristic of normal hadronic amplitudes.  Following the model
of Ref.~\refcite{BHPRL90}, we consider a standard Reggeon at
$\alpha_R=\frac12$, an Odderon exchange term, a pseudoscalar
exchange term, and a term at $\alpha_R= -1$, in addition to the
Pomeron-exchange term.

The Pomeron exchange has the intercept $\alpha_P=1+\delta$.  For
the amputated $\bar{q}-N$  amplitude $T_{\bar{q}N}$ and $q-N$
amplitude $T_{qN}$  with $q=u$, and $d$, $N=p$, and $n$, we assume
the following parameterization, including terms which represent
pseudoscalar Reggeon exchange.  The resulting amplitude is:
\begin{eqnarray}
T_{\bar{u}-p} &=& \sigma \Bigg[ s \left(i + \tan
\frac{\pi\delta}{2}\right) \beta_1 (\tau^2) - s \beta_{\cal O}
(\tau^2) - (1-i) s^{1/2} \beta_{1/2}^{0^+}(\tau^2)\\ \nonumber
&\quad& + (1+i) s^{1/2} \beta_{1/2}^{0^-}(\tau^2) - (1-i) s^{1/2}
\beta_{1/2}^{1^+}(\tau^2) + W (1-i) s^{1/2} \beta_{1/2}^{\rm
pseudo}(\tau^2) \\[1ex] &\quad&  + (1+i) s^{1/2}
\beta_{1/2}^{1^-}(\tau^2) + i s^{-1} \beta_{-1}^u(\tau^2)\Bigg].
\nonumber \label{Tu}
\end{eqnarray}
Other quark amplitudes are obtained from this one. The isovector
piece changes sign going to the $d$ quark amplitude, and the odd C
terms also change sign when one deals with the corresponding
antiparticle amplitude. We also include strange quarks, with
similar expressions. Although the phases of all these Regge
exchanges are fixed, their strengths are taken as parameters,
whose values are obtained from proton and neutron structure
functions data and known quark distribution parameterizations,
which in the model are related to the above amplitudes through:
\begin{eqnarray}
xq^{N_0}(x)&=&\frac {2} {(2\pi)^3} \frac{C x^2}{1-x} \int ds d^2
{\bf{k}}_{\bot} \rm{Im} T_{N_0}(s,\mu^2).
\end{eqnarray}

\section{Nuclear shadowing and antishadowing effects due to multiple
scattering}

Now let us turn to the scattering on a nuclear ($A$) target. The
$\bar{q}-A$ scattering amplitude can be obtained from the
$\bar{q}-N$ amplitude according to Glauber's theory as follows,
\begin{equation}
T_{\bar{q}A}=\sum_{k_1=0}^{Z} \sum_{k_2=0}^{N} \frac{1}{k_1+k_2}
\left (
\begin{array}{c}
Z+N\\ k_1+k_2
\end{array}
\right ) \frac{1}{M} \alpha^{k_1+k_2-1} \left (T_{\bar{q}p}\right
)^{k_1} \left (T_{\bar{q}n}\right )^{k_2}
 \theta (k_1 + k_2 -1)
\end{equation}
where $M=\rm{Min}\{k_1+k_2, Z\}- \rm{Max}\{k_1+k_2-N, 0\} +1$ and
$\alpha=i/({4 \pi p_{c.m.} s^{1/2} (R^2 + 2 b)})$, with
$R^2=\frac23 R_0^2, R_0=1.123 A^{1/3} \rm{fm}$, and
$b=10~(\rm{GeV/c})^{-2}$. Then the nuclear quark distributions are
readily obtained, and we get predictions for the ratio of
structure functions $F_2^A/F_2^N$, which gives an excellent
description of the experimental data~\cite{Sha1,Sha2}, showing
both shadowing and antishadowing. Furthermore, it agrees
remarkably well with the data for the ratio
$F_{2A}^{neutrino}/F_{2A}^{muon}$. For details see
Ref.~\refcite{BSY}.

We can now apply this same nuclear distribution functions to
neutrino deep inelastic scattering in nuclei. Our results show
that shadowing is similar for electromagnetic and weak currents,
but that there is a much stronger antishadowing effect for
antineutrinos, and that neutral and charge currents give different
antishadowing. Since in our nucleon quark distributions
parametrization there is still some freedom, especially in the
strange quark case, the details of these results could change, but
certainly not the overall picture, which shows a substantially
different antishadowing for charged and neutral current reactions.



\section{Nuclear effects on extraction of $\sin^2\theta_W$}

The observables measured in neutrino DIS experiments are the
ratios of neutral current to charged current events; these are
related via Monte Carlo simulations to $\sin^2\theta_W$. In order
to examine the possible impact of nuclear shadowing and
antishadowing corrections on the extraction of $\sin^2\theta_W$,
we will consider the Paschos and Wolfenstein relation~\cite{PW}
\begin{eqnarray}
\label{RNMOD}
R_N^{^-} = \frac{\sigma (\nu_\mu + N \to \nu_\mu +
X) - \sigma (\bar\nu_{\mu} + N \to \bar\nu_{\mu} + X)} {\sigma
(\nu_{\mu} + N \to \mu^- + X) - \sigma (\bar\nu_{\mu} + N \to
\mu^+ + X)} = \rho_0^2 \left(\frac{1}{2} - \sin^2\theta_W\right),
\end{eqnarray}
and a similar expression for the case in which there is a nuclear
target ($N \rightarrow A$). We estimate the nuclear effects on the
weak mixing angle in the following way. First, we use the cross
sections to calculate the Paschos-Wolfenstein ratios
$R_A^{^-}(\sin^2\theta_W)$ and $R_N^{^-}(\sin^2\theta_W)$ for
various values of $\sin^2\theta_W$, and using the NuTeV cutoffs.
Second, we extract $\rho^2$ by means of Eq.~(\ref{RNMOD}).  We
find a weak dependence of $\rho^2$ on $\sin^2\theta_W$ and
$\rho^2$ very close to $1$. Finally, we use the obtained $\rho^2$
to extract the shadowing/antishadowing effect on the weak-mixing
angle $\Delta \sin^2\theta_W$ from the modified relation:
\begin{eqnarray}
R_A^{^-}(\sin^2\theta_W) = \rho^2 \left(\frac{1}{2} -
(\sin^2\theta_W +\Delta \sin^2\theta_W) \right). \label{RAMOD}
\end{eqnarray}
We have have found that the nuclear modification to the
weak-mixing angle is approximately $\delta \sin^2\theta_W=0.001.$
The value of $\sin^2\theta_W$ determined from the NuTeV
experiment, without including nuclear shadowing/antishadowing due
to multiple scattering, is in absolute value $ 0.005$  larger than
the best value obtain from other experiments. The model used here
to compute nuclear shadowing/antishadowing effect would reduce the
discrepancy between the  neutrino and electromagnetic measurements
of $\sin^2\theta_W$ by about $20\% $, although this could be made
larger by modifying the strange quark distribution, which at
present is not very well known.




\section*{Acknowledgements}

Work done in collaboration with S. J. Brodsky and J. J. Yang, and
partially supported by Fondecyt (Chile) under grant 1030355.


%
%
%
%


\end{document}